\documentstyle[12pt]{article}
\newcommand{\slp}{\raise.15ex\hbox{$/$}\kern-.57em\hbox{$\partial$}}
\newcommand{\sla}{\raise.15ex\hbox{$/$}\kern-.57em\hbox{$a$}}
\newcommand{\slA}{\raise.15ex\hbox{$/$}\kern-.57em\hbox{$A$}}
\newcommand{\slb}{\raise.15ex\hbox{$/$}\kern-.57em\hbox{$b$}}
\newcommand{\be}{\begin{equation}}
\newcommand{\ee}{\end{equation}}
\newcommand{\bear}{\begin{eqnarray}}
\newcommand{\ear}{\end{eqnarray}}
\newcommand{\ba}{\begin{eqnarray*}}
\newcommand{\ea}{\end{eqnarray*}}

\begin{document}
\begin{titlepage}
\setcounter{page}{1}
\begin{flushright}
HD--THEP--00--10\\
\end{flushright}
\vskip1.5cm
\begin{center}
{\large{\bf Thermal Quantum Fields in Static Electromagnetic
Backgrounds}}

\vspace{1cm}
A.A. Actor \\
\small{\it Department of Physics, The Pennsylvania State University}\\
\small{\it Lehigh Valley Campus, Fogelsville, PA 18051, U.S.A.}
%%%%%%%%
\footnote{e-mail: aaa2@psu.edu}
%%%%%%%%%
\\
K. D. Rothe
%%%%%%%%
\footnote{e-mail: k.rothe@thpys.uni-heidelberg.de}
%%%%%%%%
\\
\small{\it Institut  f\"ur Theoretische Physik - Universit\"at Heidelberg}\\
\small{\it Philosophenweg 16, D-69120 Heidelberg}
\date{\today}
\end{center}

\begin{abstract}
{\small We present and discuss, at a general level, new mathematical results
on the spatial nonuniformity of thermal quantum fields coupled minimally
to static background electromagnetic potentials. Two distinct examples
are worked through in some detail: uniform (parallel and perpendicular) background electric and
magnetic fields  coupled to a thermal
quantum scalar field.} 
\end{abstract}
\end{titlepage}

%%%%%%%%%%%%%%%%%%%%%%%%%%%%%%%%%%%%%%%%%%%%%%%%%%%%%%%%%%%%%%%%%%%%%%%%%%%%
\section{Introduction and General Results}

In thermal quantum field theory (hereafter thermal QFT) with matter fields 
coupled to an external gauge potential it has long
been known \cite{1}-\cite{7} that a \underbar{constant} Euclidean gauge 
potential
%%%%%%%%%%%%%
\footnote{Throughout this paper we
shall move freely back and forth between
Euclidean and Minkowski spacetime. 
For sake of clarity we attach the label
E to (certain) Euclidean quantities, and
in particular to the Euclidean gauge potential.}
%%%%%%%%%%%%
$A^E_0=a=constant$ is a compact physical parameter.
Upon first encounter this may seem strange, because at zero
temperature $(T=0)$ one can simply remove such a constant by
means of a gauge transformation $A^E_\mu\to A^E_\mu-\partial_\mu\lambda$
with $\lambda=x_0A^E_0$. However, in the Matsubara or imaginary time
formalism \cite{8,9} (which we use throughout this paper) bosonic (fermionic) matter fields must be periodic (antiperiodic) functions of (Euclidean)
time, with period $\beta = {1\over T}$. This restricts the allowed gauge transformations to
those satisfying $\lambda(x_0+\beta)=\lambda(x_0)+\eta\pi\ mod\; 2\pi$, 
with $\eta=0$ \cite{1} for bosonic (fermionic) matter fields. Hence
the constant 
$a$ can only be removed if $a=(2N+\eta)
(\pi/\beta)$. 

It follows from above that it is always possible to gauge 
an arbitrary $A_0^E(\vec x)$ \underbar{into} the
physical interval
$0\leq A^E_0(\vec x)\leq \frac{2\pi}{\beta}$ at every point $\vec x$.
Moreover, $\beta A^E_0=2\pi$ is gauge-equivalent
to $\beta A^E_0=0$; hence the real dimensionless quantity $\beta A^E_0$ 
is compact -- an angle.
What physics underlies this angular variable? 
A specific
and complete answer to this question can be given. Though the essential 
facts have been known for a long time, many in the
field theory community seem not to be aware of this physical picture, 
as we have not found it discussed in the literature.

The physical meaning of the angle $\beta A^E_0$ emerges from the old
observation 
%%%%%%%%%%%
\footnote{Note that we absorb
the unit $e$ of electric charge carried by the particles of 
the matter field into the gauge potential $A^E_\mu$. Thus the
covariant derivative is $D_\mu=\partial_\mu-iA^E_\mu$, and
$A^E_\mu$ has, in natural units, the dimension of energy. In
particular, $A^E_0(\vec x)$ is the electrostatic energy
of a positive charge at point $\vec x$. To discuss voltage
explicitly, we refer to $A^E_0(\vec x)/e$.}
%%%%%%%%%%
(see e.g. refs. \cite{10, 11}) that an imaginary
constant Euclidean gauge potential
\be\label{1.2}
A^E_0=iA_0,\quad A_0=\mu\ee
corresponds to a chemical potential $\mu$ for the thermal matter 
field to which $A_\mu$ is minimally coupled. This is true for
both spinor and scalar thermal fields. Eq. (\ref{1.2}) is
written to emphasize that, eventually, one must change from
the background Euclidean potential $A_\mu^E = (A^E_0,\vec A)$ to the corresponding
background Minkowski potential $A_\mu=(-iA^E_0, \vec A^E)$ when
writing down one's final physical formulae. The identification
$A_0=\mu$ as a chemical potential is merely the recognition that
for a charged particle in a uniform background voltage $A_0/e$
pervading all of space, $A_0$ has the meaning of a chemical
potential: it is the electrostatic energy that must be expended
to create the charged particle at whatever position $\vec x$ this
particle occupies. In sect. 2 we similarly recognize that a 
nonuniform background voltage $A_0(\vec x)/e$ has locally
this same interpretation.

The physical significance of a constant
$A^E_0/e$ is now clear. It represents,
in Euclidean language, a uniform voltage
$A_0/e$ throughout space. Clearly
$A_0=constant$ is a true physical parameter -- noncompact
obviously -- a voltage which is felt by any real charged particle,
and felt in particular by the real particles
of the thermal plasma. In the limit
$T\to 0$ this plasma disappears, leaving the
virtual particle sea, which is \underbar{not} sensitive
to a uniform voltage throughout space. The 
sea knows nothing about uniform background voltage because the
virtual pairs of which the sea consists have precisely zero
electrostatic energy in such a background. 

Once it is known that Euclidean $\beta A^E_0$ is an angular
variable one can, using the power of Fourier analysis, write
all gauge-invariant physical functions as Fourier cosine
series in this angle. Thus for, respectively, the diagonal elements of the
Euclidean heat kernel $h^{(\beta)}$, effective Lagrangian ${\cal L}^{(\beta)}$ and stress tensor
$T^{(\beta)}_{\mu\nu}$ of the thermal quantum field we can write
\be\label{1.3}
h^{(\beta)}(t|\vec x,\vec x)=\sum^\infty_{n=0}(\pm)^nh^{(\beta)}_n(t|\vec x)
\cos (n\beta A^E_0(\vec x)),\ee
\be\label{1.4}
{\cal L}^{(\beta)}(\vec x)=\sum^\infty_{n=0}(\pm)^n
{\cal L}^{(\beta)}_n(\vec x)\cos(n\beta A^E_0(\vec x)),
\ee
\be\label{1.5}
T^{(\beta)}_{\mu\nu}(\vec x)=\sum^\infty_{n=0}(\pm)^nT^{(\beta)}_{\mu\nu;n}
(\vec x)\cos (n\beta A^E_0(\vec x)).\ee
Here in each formula the coefficients of $\cos (n\beta A^E_0(\vec x))$ 
depend on $\vec{\cal E}(\vec x)=-\vec\nabla A^E_0(\vec x)$ 
and $\vec B(\vec x)=\vec\nabla \times\vec A(\vec x)$, but not
directly on $A^E_0(\vec x)$. Our use of $A^E_0(\vec x)$ here
implies an arbitrary static Euclidean gauge potential, and that
will be our final result. Sect. 2 is devoted to demonstrating
the above implied compactification of $A^E_0(\vec x)$ at the local level. 
The nonalternating/alternating sign $(\pm)^n$ is appropriate
for scalar/spinor thermal matter fields. One could equivalently disregard this sign 
and replace $A^E_0$ by $A^E_0-\eta(\pi/\beta)$ everywhere
with $\eta=0$ \cite{1} for scalar (spinor) field.

It is worth pointing out that eqs. (\ref{1.3})-(\ref{1.5})
display the expected complete separation of all functions
characterizing the thermal field into parts representing the virtual 
sea and thermal plasma. E. g. for the heat kernel we have
\be\label{1.6}
h^{(\beta)}=h_{sea}+h^{(\beta)}_{plasma}\ee
where
\be\label{1.7}
h_{sea}= h_0(t|\vec x),
\ee
\be\label{1.8}
h^{(\beta)}_{plasma}=\sum^\infty_{n=1}(\pm)^n h^{(\beta)}_n(t|\vec x)
\cos (n\beta A^E_0(\vec x)).
\ee
where $h_0(t|\vec x)$ is the $T=0$ heat kernel and represents
the vacuum or virtual
particle sea contribution to the thermal heat kernel. The vacuum is 
always independent of $T$: its virtual particles do not have
the prolonged existence needed to come into thermal equilibrium
with anything. Neither does the vacuum feel directly an applied
voltage, so $h_{sea}$ cannot depend explicitly on $A^E_0$. Similarly 
${\cal L}^{(\beta)}$ and $T^{(\beta)}_{\mu\nu}$ separate into sea $(n=0)$ and
plasma $(n>0)$ parts, the
former being independent of $\beta$ and $A_0^E$, while the latter 
depend on both $\beta$ and $A^E_0$.

Before embarking on calculations, a few words about background $\vec {\cal E}$
and $\vec B$ fields interacting with the vacuum and with the thermal
plasma may be of use to some readers.

\bigskip\noindent 
{\it Virtual sea}

\medskip
The standard visualization of vacuum quantum fluctuations as virtual
pairs -- initially zero-length ``vacuum dipoles'' which grow to maximum size,
then shrink again to zero length and annihilate away --
enables one also to visualize the effect of a background $A_0$, as well as 
the effect of electric and magnetic fields, on these fluctuations. Due to their mutual
``binding'', virtual pairs do not feel $A_0(\vec x)$ directly.
This was already mentioned for constant $A_0$. 
The vacuum pair however couples to any nonuniformity in $A_0
(\vec x)$ -- i.e. to the electric field.

A background electric field exerts an aligning torque on vacuum dipoles
-- the famous ``vacuum polarization'' effect. $\vec E$ also tries
to stretch or shorten these nonrigid dipoles, depending on their
orientation relative to $\vec E$. A vacuum dipole whose moment is
parallel to $\vec E$ will be stretched and, perhaps, even given
enough external energy to break apart into real particles. Dipoles
antiparallel to $\vec E$ will be shortened; those perpendicular
to $\vec E$ only rotated. Real pair creation occurs from the vacuum,
preferentially along the direction of $\vec E$, but not perpendicular
to $\vec E$. Pair creation is independent of $T$. As in Schwinger's
original $T=0$ calculation \cite{12} and the subsequent literature
known to us (see e.g. the books \cite{13}-\cite{15}), our calculations
predict the phenomenon of pair creation, but do not take account of these
pairs once they have been produced. All such calculations treat pair
creation as a perturbation of a pre-existing many-body system --
the virtual sea or the sea plus plasma, with fixed background
$\vec E$ and $\vec B$ -- whose subsequent development 
is not investigated. 
\footnote{For sake of completeness we mention here the non-equilibrium approach to background fields \cite{27}. In this approach the existence of an electric field $\vec E_0$ is assumed as initial condition with, say, no real pairs present. Pair production ensues, and the produced pairs serve in turn as sources of an additional electric field: charge separation occurs, first partially, cancelling, then strongly overcancelling $\vec E_0$. More pairs are produced and things go in reverse. Eventually plasma oscillations set in.
These calculations involve large sets of coupled equations and are intensively numerical. There is no thermal equilibrium, and hence no temperature. Our analytic work here could serve as the $T>0$ initial conditions for such numerical investigations.}

A background $\vec B$ cannot transfer energy to individual particles, and
therefore cannot cause particle production from the vacuum. Evidently,
$\vec B$ acts locally  (more or less rigidly) to displace 
vacuum pairs, but not to stretch or shorten them. To this limited
extent the vacuum can be aware of $\vec B$.

\bigskip\noindent 
{\it Thermal plasma}

\medskip
The thermal plasma is a neutral quantum gas of unbound charged
real particles. Pair creation
does not occur from the thermal plasma. Through their electrostatic energy these particles
know individually about the background potential $A_0(\vec x)$ --
our main point leading to eqs.(\ref{1.3})-(\ref{1.5}). 
If $\vec E(\vec x)=-\vec \nabla A_0(\vec x)\not=0$, these particles
feel individually the Coulomb force $\vec F=q\vec E$, which of course 
accelerates $q= e(-e)$ parallel (antiparallel) to $\vec E$. Neither 
our calculation, nor others in the $T>0$ literature follow up the consequences
of this acceleration. Through their thermal motion,
the plasma particles also feel the magnetic force $\vec F=q\vec v\times\vec B$ perpendicular to $\vec B$. Functions describing the thermal plasma
therefore depend on $A_0(\vec x), \vec E$ and $\vec B$. 

%%%%%%%%%%%%%%%%%%%%%%%%%%%%%
\section{Euclidean Spacetime}
\setcounter{equation}{0}
\subsection{Compactification}
%%%%%%%%%%%%%%%%%%%%%%%%%%%%%
The Fourier series (\ref{1.3})-(\ref{1.5}) rest upon the fact that
in any gauge-invariant quantity, $\beta A^E_0(\vec x)$ plays the role of
a \underbar{local} compact angle,
$0\leq\beta A^E_0(\vec x)\leq 2\pi$, since, as we have seen, 
one can gauge this function into the interval $[0,\frac{2\pi}{\beta}]$ at any point $\vec x$ in space.
Moreover, the upper and lower ends of this interval can be identified
by a gauge transformation. Note that this compactification does not extend to a time-dependent
gauge potential $A^E_0$. Indeed, within the
Matsubara formalism one cannot accommodate time-dependent
backgrounds of any kind.

The Fourier series (\ref{1.3})-(\ref{1.5}) express, or are the 
result of, a remarkable series resummation, as we shall
illustrate in subsection 2.3 below. Let us briefly
recall the mode-sum construction of the thermal heat kernel
for a scalar field coupled to an arbitrary static
background potential $A_\mu(\vec x)$ \cite{16}. The spacetime
Matsubara modes
\be\label{2.4}
\phi_{mp}(x_0,\vec x)=\frac{1}{\sqrt\beta}e^{i(2\pi m/\beta)x_0}\varphi_{mp}(\vec x)
\ee
satisfy  $(-D^2) \phi_{mp}=\lambda^2_{mp}\phi_{mp}$,
where $D_\mu = \partial_\mu - iA^E_\mu$, $m$ runs over all integers, and $p$ is a collective
label for all spatial directions. The modes $\varphi_{mp}
(\vec x)$ satisfy
\be\label{2.5}
\left[(A_0^E(\vec x)-\frac{2\pi m}{\beta})^2-(\vec\nabla-i\vec A(\vec x)
)^2\right]\varphi_{mp}(\vec x)=\lambda^2_{mp}
\varphi_{mp}(\vec x).\ee
Here the $(A^E_0-2\pi m/\beta)^2$ term in the bracket 
has particular importance. It couples position $\vec x$ to
the Matsubara label $m$. Also, it displays the local compactification
of $A^E_0$: the shift $A^E_0\to A_0^E-2\pi
N/\beta$ merely shifts the Matsubara label, $m\to m+N$.

The thermal heat kernel of the operator $-D_\mu^2$ is defined by
\be\label{2.6}
h^{(\beta)}(t|x,y)=\sum_{m,p}e^{-t\lambda^2_{mp}}\phi_{mp}(x)\phi
^*_{mp}(y).\ee
The corresponding diagonal elements 
\bear\label{2.7}
h^{(\beta)}t|x,x)&=&\sum_{m,p}e^{-t\lambda^2_{mp}}
|\varphi_{mp}
(\vec x)|^2\nonumber\\
&=&\sum_n h^{(\beta)}_n(t|\vec x)\cos (n\beta A^E_0(\vec x))\ear
display the mode sum resummation to a Fourier series, alluded to
above. For reasons of gauge invariance
the coefficients in eq. (\ref{2.7}) can only depend on the Euclidean electric
field $-\vec \nabla A^E_0$ and  magnetic field $\vec B=\vec \nabla
\times \vec A$, but not on the potential $A^E_0$ directly.
Except for the $n=0$ coefficient, they also depend on the temperature. 
In the limit $T\to 0$ all the $h^{(\beta)}_{n\not=0}
$ vanish exponentially (the thermal plasma disappears), and what
remains is the $T=0$ heat kernel (\ref{1.7}) for the virtual
sea. All of these statements are illustrated by the example 
in subsection 2.3 below, and those in sections 4 and 5.

%%%%%%%%%%%%%%%%%%%%%%%%%%%%%%%%%%
\subsection{Effective Lagrangians}
%%%%%%%%%%%%%%%%%%%%%%%%%%%%%%%%%%

Much of the early work on thermal quantum fields coupled to background
gauge fields was concerned with effective Lagrangians for the potential
$A^E_0$ (see e.g. refs. \cite{2}, \cite{6}, \cite{7}). A related
theme was the study of ``order parameters'' which signaled the
deconfinement phase transition at high $T$ in nonabelian gauge
theories (see e.g. refs. \cite{3, 4, 5}). Our formula (\ref{1.4}) has a
natural interpretation as an effective Lagrangian $={\cal L}^{(\beta)}$. 
The coefficients ${\cal L}_n^{(\beta)}(\vec x)$ of $\cos (n \beta A^E_0)$ in
eq. (\ref{1.4}) are actually functions of $\vec{\cal E}
\cdot\vec{\cal E}=(\vec\nabla A^E_0)^2$
(not to mention $\vec B\cdot\vec B$ which we suppress 
here), and therefore play the role of (very complicated) ``kinetic
terms'' in ${\cal L}^{(\beta)}(A^E_0)$. The $\cos (n \beta A^E_0)$ 
factors play the role of ``potential terms'' in the same Lagrangian.
An expansion of ${\cal L}^{(\beta)}$ in powers of $\vec{\cal E}
\cdot\vec{\cal E}$ and $(A^E_0)^2$ has the form
\bear\label{2.8}
{\cal L}_\beta&=&a_0+a_1(A^E_0)^2+\cdot\cdot\cdot\nonumber\\
&&+(\vec\nabla A^E_0)^2[b_0+b_1(A^E_0)^2+\cdot\cdot\cdot]\nonumber\\
&&+\cdot\cdot\cdot\ear
where we find the conventional kinetic term among all the
others. Here the coefficients $a_n,b_n,...$ are independent
of $A^E_0$ and $\vec {\cal E}$ (but depend on background $\vec B$).

%%%%%%%%%%%%%%%%%%%%%%%%%%%%%%%%%%%%%%%%%%%
\subsection{Fourier series and resummation}
%%%%%%%%%%%%%%%%%%%%%%%%%%%%%%%%%%%%%%%%%%%

In the following we wish to illustrate, for the case
of fermions in one space dimension, how the Fourier series in (\ref{1.2}) is the result of an infinite resummation of the Matsubara sum.
Thermal fermionic fields must, of course,
satisfy antiperiodic boundary condition in $x_0$. Let us consider eqs. (\ref{2.4}), (\ref{2.5}) for a scalar field
satisfying the \underbar{antiperiodic} boundary condition
$\phi(x_0+\beta)=-\phi(x_0)$ in Euclidean time. 
This only requires the replacement $m\to m+1/2$
in eqs. (\ref{2.4}), (\ref{2.5}). In the mode
equation (\ref{2.5}) the $1/2$ can be absorbed into the gauge
potential, $A^E_0\to A^E_0-\pi/\beta$, leaving everything else
just as it was. Consequently, the only change in the heat kernel
(\ref{2.7}) and related Fourier series
is
\[\cos (n\beta A^E_0)\to (-)^n
\cos (n\beta A^E_0).\]

The preceding argument indicates
that for thermal Fermi fields one will have the alternating 
signs displayed in eqs. (\ref{1.3})-(\ref{1.5}).

As is well known (see e.g. ref. \cite{17,book} and
references therein),
the small $t$ expansion of heat kernels is of the form
\be\label{2.9}
h^{(\beta)}(t;x,x)\sim\sum_{k=0}^\infty t^{(k-d-1)/2}a^{(\beta)}_k(\vec x)\;,
t\to 0
\ee
where $d$ is the space-time dimension, and where the coefficients $a_k^{(\beta)}(\vec x)$ depend on 
the quantum field as well as the structure of space-time in
which the quantum field lives. Given our knowledge of the Fourier
series (\ref{2.7}), we can make the obvious prediction

\be\label{2.10}
a^{(\beta)}_k(\vec x)=\sum_n(\pm)^na^{(\beta)}_{kn}(\vec x)\cos (n\beta A^E
_0(\vec x)).\ee
This statement goes far beyond the standard lore of asymptotic
heat kernel expansions. It is instructive to see how this periodicity
comes about in the context of a Seeley expansion.

In \cite{18} it was shown that in two space-time dimensions the heat kernel for the Dirac operator of massless fermions
in an external, static gauge field $A_0^E = ({\cal E}x_1 + \frac{2\pi a}{\beta}, A_1=0)$ takes the form
\be\label{diagonallocalheatkernel2}
tr h^{(\beta)}(t;x,x) = \frac{\cal E}{2\pi}
\left(\frac{1}{\tanh {\cal E} t}\right)
\left\{1 + 2\sum_{n=1}^{\infty} (-1)^n 
\cos\left[n\beta ({\cal E}x_1 + \frac{2\pi}{\beta} a)\right]
e^{-\frac{n^2\beta^2 {\cal E}}{4 \tanh {\cal E} t}}\right\}\;,
\ee
where we see the anticipated alternating sign in the sum.
Expanding the ${\cal E}$-dependent multiplicative factor as well as the exponential in powers of $t$,
\[
\frac{{\cal E}}{2\pi}
\left(\frac{1}{\tanh {\cal E} t}\right)= 1 + \frac{1}{3}({\cal E}t)^2 + \cdot\cdot\cdot
\]
\[
e^{-\frac{n^2\beta^2 {\cal E}}{4 \tanh {\cal E} t}} =  (1-\frac{1}{12}n^2\beta^2{\cal E}^2 t + \cdot\cdot\cdot)e^{-\frac{n^2\beta^2}{4t}}
\]
one finds from (2.8) for the diagonal elements of the heat kernel,
\bear\label{asymptexp}
2\pi tr h^{(\beta)}(t;x,x) &=& \frac{1}{t}\left[ 1 + 2\sum_{n=1}^\infty (-)^n e^{-\frac{n^2\beta^2}{4t}}\cos[n({\cal E}\beta x_1 + 2\pi a]\right]\\
&-& \frac{1}{6}\sum_{n=1}^\infty(-)^n e^{-\frac{n^2\beta^2}{4t}}
n^2\beta^2{\cal E}^2\cos[n({\cal E}\beta x_1 + 2\pi a)] + O(t)\;.\nonumber
\ear

On the other hand it was shown in Ref. \cite{Natividade} (see also 
\cite{18}) that the above heat kernel possesses a formal
expansion of the form
\be\label{Seeleyexpansion1}
h^{(\beta)}(t;x,x) = \frac{1}{4\pi t}\sum_{\ell=0}^\infty
a_\ell(x;\frac{\sqrt t}{\beta})t^\ell
\ee
where
\bear\label{Seeleycoefficient1}
a_\ell\left(x;{\sqrt t\over\beta}\right)&=& \sqrt{{4\pi t}\over\beta^2}
\int {dk_1\over\sqrt\pi}\sum_{m=-\infty}^\infty
\,e^{-(k_1^2+\bar\omega_m^2\left({\sqrt t 
\over\beta}\right))} \\
&\times& \left\{  \sum_{r=0}^\ell
\sum_{dist.perm.}\frac{(-1)^{\ell-r}}{(\ell+r)!}(2ik\cdot 
{D})^{2r} \hat D^{\ell-r}\right\}_{k_2 =
\bar\omega_m\left({\sqrt t\over\beta}\right)}\cdot {\bf 1}\nonumber
\ear
where the sum is over all distinct permutations, $\bar\omega_m$  are the scaled Matsubara frequencies
\[
\bar\omega_m\left({\sqrt t\over \beta }\right)= 2\pi\left(m+{1\over 2}\right)\left(
{\sqrt t\over \beta}\right) = \sqrt t \omega_m \quad. 
\]
and
\[
D_\mu = \partial_\mu - iA^E_\mu \;,\quad \hat D = -\partial^2 + (A^E_0)^2 + X
\]
with $X$ a matrix valued field ($\epsilon_{01}=1$)
\[
X = -\frac{1}{2}\gamma^5 \epsilon_{\mu\nu}\partial_\mu A^E_\nu\;.
\]
In (\ref{Seeleycoefficient1}) we have already taken account of the fact, that only even powers in $k_1$ and $\bar\omega_m$ will contribute to
the integral and sum in  
(\ref{Seeleycoefficient1}). The leading contribution to (\ref{Seeleycoefficient1}) for $t\to 0$
is given by the $r = \ell$ term in the sum, and in particular 
by the term $\bar\omega_m A^E_0$ in
$ik\cdot D$. Hence,
\bear\label{Seeleycoefficient2}
a_\ell\left(x;{\sqrt t\over\beta}\right)&\approx&\sqrt{\frac{t}{\beta^2}}
\int \frac{dk_1}{\sqrt{\pi}} e^{-k_1^2}\nonumber\\
&\times& \sqrt{4\pi} \frac{1}{(2\ell) !} \sum_m e^{-\bar \omega^2_m}(4\omega_m^2)^\ell (A_0^E)^{2\ell}\nonumber\\
 &=& \sqrt{ t\over{\beta^2}}
\bar I_\ell \frac{(A_0^E)^{2\ell}}{(2\ell)!}  + O(\frac{t}{t^\ell})\quad (\ell>0)\;.
\ear
where
\be
\bar I_\ell = \sqrt{4\pi} \sum_{m=-\infty}^\infty (4\bar\omega_m^2)^\ell e^{-\bar\omega^2_m}\;. 
\ee
We obtain the expansion of $\bar I_\ell$ in powers of $t$ by repeatedly differentiating the Jacobi identity
\be\label{generalizedJacobi}
\sum_{m=-\infty}^\infty {e}^{-\tau\left[2\pi\left(m+{1\over 2}
\right)\right]^2} = \sqrt{\frac{1}{4\pi\tau}}\left[ 1+ 2\sum_{n=1}^\infty (-)^n 
{ e}^{-\frac{n^2}{4\tau}}\right]\quad.\label{n4.15}
\ee
with respect to $\tau$,and setting $\tau = \frac{t}{\beta^2}$.
We thus find
\[
\bar I_0 = \sqrt{\frac{\beta^2}{t}}[1 + 2\sum_{n=1}^\infty (-)^ne^{-\frac{n^2\beta^2}{4t}}]\;,
\]
\[
\bar I_\ell = \sqrt{\frac{\beta^2}{t}}
2(-1)^\ell\sum_{n=1}^\infty(-)^ne^{-\frac{n^2\beta^2}{4t}}
\left[\left(\frac{n^2\beta^2}{t}\right)^\ell + O(\frac{t}{t^\ell}) \right]\;.
\]
We thus finally have from (\ref{Seeleycoefficient2})
\bear\label{leadingSeeleycoefficient}
a_0\left(x;\frac{\sqrt t}{\beta}\right) &=& \left[ 1+ 2\sum_{n=1}^\infty (-)^n 
e^{-{n^2\beta^2\over 4 t}}\right]\;, \\
t^\ell a_\ell\left(x;\frac{\sqrt{t}}{\beta}\right) &=& 2(-)^\ell \sum_{n=1}^\infty (-)^n 
e^{-{n^2\beta^2\over 4 t}}\left[\frac{(n\beta {A^E_0})^{2\ell}}{(2\ell)!} + O(t)\right]
 \quad (\ell > 0)\;.
\nonumber
\ear
Hence
\be\label{sumSeeley}
\sum_{\ell} t^\ell a_\ell\left(x;\frac{\sqrt t}{\beta}\right)=
1 + 2\sum_{n=1}^{\infty}(-)^n e^{-{n^2\beta^2\over 4 t}}
\left[\cos(n\beta A^E_0(x))+ O(t)\right]\;.
\ee 
Substitution of (\ref{sumSeeley}) into (\ref{Seeleyexpansion1})
 reproduces the leading term
in the small $t$ expansion (\ref{asymptexp}) of the heat kernel.

The corresponding calculation of next to leading order is very cumbersome
due to the non-commutativity of the operators appearing in the expansion (\ref{Seeleycoefficient1}), and we shall not persue this any further.

%%%%%%%%%%%%%%%%%%%%%%%%%%%%%
\section{Minkowski Space-time}
\setcounter{equation}{0}
%%%%%%%%%%%%%%%%%%%%%%%%%%%%%
Once we know eqs.(\ref{1.3})-(\ref{1.5}) are valid for an arbitrary
static Euclidean background potential $A^E_0(\vec x)$, it is clear
that we must continue these formulae to Minkowski space-time in
order to make them physically meaningful. Thermal
equilibrium having been assumed, there is no $x_0$ dependence
anywhere to deal with. The only continuation needed is in the
gauge potential
\be\label{3.1}
A^E_0(\vec x)\to iA_0(\vec x)\ee
and correspondingly in the background electric field
\be\label{3.2}
\vec{\cal E}=-\vec\nabla A^E_0 \to i\vec E=-i\vec\nabla A_0.\ee
Making the change (\ref{3.2}) wherever $\vec{\cal E}$ appears in
the coefficients in eqs. (\ref{1.3})-(\ref{1.5}) as well as the replacement
$\cos n\beta A^E_0 \to \cosh n \beta A_0$,
eqs.(\ref{1.3})-(\ref{1.5}) become Minkowski space-time statements.
The latter are the central results of the present article, obtained
by general arguments based on gauge invariance and Fourier analysis.

At this point examples may be helpful. Let us quote the following
two thermal heat kernels from refs. \cite{16,18} where the detailed
calculations can be found. 

Continuing (\ref{diagonallocalheatkernel2}) to Minkowski space we have for a
spinor field in 1 spatial dimension coupled to
$A_\mu=(Ex_1+\mu,0)$\cite{18}:
\be\label{3.4}
tr h^{(\beta)}(t|x,x)=\frac{E}{2\pi \tan Et}\sum^\infty_{n=-\infty}(-)
^ne^{-n^2\beta^2E/4 \tan Et}e^{n\beta(Ex_1+\mu)}.\ee

For a scalar field in $d$ spatial dimensions coupled to $A_\mu=
(Ex_1+\mu,\vec 0)$ \cite{16}:
\be\label{3.5}
h^{(\beta)}(t|x,x)=(4\pi t)^{-\frac{d-1}{2}}\frac{E}{4\pi \sin Et}
\sum^\infty_{n=-\infty}e^{-n^2\beta^2E/4\tan Et}
e^{n\beta(Ex_1+\mu)}.\ee

The background in eqs. (\ref{3.4}), (\ref{3.5}) is a uniform electric
field in the $x_1$ direction. The vacuum $(n=0)$ contributions
\bear\label{3.6}
{\rm spinor}&:&\quad tr\ h^{(\beta)}(t)_{sea}=\frac{{E}}{2\pi \tan {E}t}\;,\nonumber\\
{\rm scalar}&:&\quad  h^{(\beta)}(t)_{sea}=(4\pi t)^{-\frac{d-1}{2}}
\frac{{E}}{4\pi \sin {E}t},
\ear
to the thermal heat kernels above display the ubiquitous singularity
at $t=0$ and, in addition, singularities at $t=q\pi/E$ with $q=1,2,3..$. 
One does not expect to find the latter singularities in a physical
heat kernel. They are present here because the vacuum is unstable:
the background electric field produces (at a temperature-independent
rate which does not directly involve the background voltage
$A_0/e$) pairs of real particles from the sea. This has been discussed
by Schwinger \cite{12} and by others (see e.g. the books \cite{13}-
\cite{15}).

The plasma contributions in eqs. (\ref{3.4}), (\ref{3.5}) -- the sum of all
$n\not = 0$ terms -- display all of the properties mentioned earlier. They are
nonsingular at $t=0$: the thermal plasma is UV-finite.
They have no singularities for $t>0$: pair production from the sea 
is temperature-independent. They vanish exponentially as $T\to 0$: the plasma disappears. Most importantly, they depend explicitly on the gauge
potential $A_0=E x_1+\mu$ in the way we expect them to.

Global studies of thermal fields coupled to a uniform background $E$
are given in refs. \cite{19}-\cite{22}. These investigations do not find
the $\cosh [m\beta(Ex_1+\mu)]$ dependence in local plasma quantities. For large $x_1$ the factors $\cosh [m\beta
(Ex_1+\mu)]$  diverge . However, the meaning of this
(apparent) divergence can be explained in very physical terms. It is 
the result of the background voltage function which is unbounded as $x_1
\to\pm \infty$, this being of course, an idealization associated with
a uniform electric field of infinite spatial extent. 
%%%%%%%%%%%%%%%%%%%%%%%%%%%%%%%%%%%%%%%%%%%%%%%%%%%%%%%%%%%%%%%%
\section{Parallel Uniform Electric and Magnetic Fields}
\setcounter{equation}{0}
%%%%%%%%%%%%%%%%%%%%%%%%%%%%%%%%%%%%%%%%%%%%%%%%%%%%%%%%%%%%%%%%%

To further illustrate the Fourier series (\ref{1.3})-(-\ref{1.5}),
we now discuss the problem of parallel uniform $\vec E$ and $\vec B$
fields coupled to a thermal scalar field. Parallel $\vec E$ and $\vec B$
exert mutually perpendicular electric and magnetic forces on individual
charged particles. Mathematically this leads to a complete
factorization of the electric and magnetic sectors. Global treatments 
of the spinor version of this problem can be found in refs. 
\cite{20}-\cite{22}.
The $T=0$ problem was solved by Schwinger long ago \cite{12}.
%%%%%%%%%%%%%%%%%%%%%%%%%%%%
\subsection{Infinite space}
%%%%%%%%%%%%%%%%%%%%%%%%%%%%

For the Euclidean gauge potential $A^E_\mu(\vec x)=({\cal E} x_1+c_0,0
,0, Bx_2+c_3)$ corresponding to a uniform background magnetic field 
$\vec B=(B,0,0)$ parallel to the Euclidean electric field 
$\vec{\cal E}=({\cal E},0,0)$, the mode equation (\ref{2.5}) separates. With 
$p=(n,n',k)$ and modes
\be\label{4.1}
\varphi_{mp}(\vec x)=\frac{1}{\sqrt{2\pi}}e^{ikx_3}\psi_{mn}
(x_1)\psi_{kn'}(x_2)\ee
eq. (\ref{2.5}) separates into
\be\label{4.2}
[-\partial_1^2+{\cal E}^2(x_1+c_0/{\cal E}-
2\pi m/\beta{\cal E})^2]\psi_{mn}(x_1)=2{\cal E}(n+1/2)
\psi_{mn}(x_1)
\ee
and
\be\label{4.3}
[-\partial^2_2+B^2(x_2+c_3/B-k/B)^2]\psi_{kn'}(x_2)
=2B(n'+1/2)\psi_{kn'}(x_2),
\ee
where $n,n'=0,1,2,...$ and
\be\label{4.4}
\lambda_{mp}^2=2{\cal E}(n+1/2)+2B(n'+1/2).\ee
A peculiarity of this spectrum is its lack of dependence on $m$ 
and $k$. This degeneracy does complicate the calculation of global
quantities but not, as we shall see, of local functions. One knows 
the eigenvalues and eigenfunctions in eqs. (\ref{4.2}),
(\ref{4.3}) since they are both harmonic oscillator (HO)
equations in $d=1$. Hence the corresponding eigenfunctions are
\be\label{2ac}
\psi_{mn}(x_1)=2^{-n/2}\frac{1}{\sqrt{n!}}\left(
\frac{{\cal E}}{\pi}\right)^{\frac{1}{4}}e^{-\frac{1}{2}{\cal E}x^2_m}
H_n(\sqrt {\cal E} x_m)\;,
\ee
with
\be\label{2ab}
x_m\equiv x_1+ \frac{c_0}{{\cal E}} -\frac{2\pi m}{\beta {\cal E}}\;,
\quad m=0,1,2,\cdot\cdot\cdot
\ee
and
\bear\label{4.5}
\psi_{kn'}(x_2)&=&2^{-n'/2}\frac{1}{\sqrt {n'!}}\left(\frac{B}{\pi}\right)^{1/4}
e^{-\frac{1}{2}B x_k^2}H_{n'}(\sqrt B x_k)
\ear
with 
\[
x_k=x_2+(c_3-k)/B\;,\quad n'=0,1,2,...\;.
\]
Here $H_n'(z)$ are Hermite polynomials satisfying $H_{n'}''-2zH_{n'}'+2nH_{n'}=0$. The (diagonal) heat kernel constructed from the modes (\ref{2ac}) is respectivley (see \cite{16}
for details)
\bear\label{h1}
h_1^{(\beta)}(t|x,x)=\frac{{\cal E}}
{4\pi \sin {\cal E}t}
\sum^\infty_{n=-\infty}e^{-n^2\beta^2{\cal E}/4\tan {\cal E}t}
e^{n\beta({\cal E}x_1+\mu)}.
\ear
The (off diagonal) heat kernel constructed from the modes (\ref{4.5}) is
\be\label{4.6}
h_2(t|x,y)=\left[\frac{B}{2\pi\sinh 2B t}\right]^{\frac{1}{2}}
\frac{1}{2\pi}\int_{-\infty}^\infty dk e^{-\frac{1}{2}B(x_k-y_k)^2\coth 2B t} e^{-Bx_k y_k \tanh Bt}\;,
\ee
where again the details of the calculation parallel those in ref. \cite{16}.

Putting things together, the diagonal heat kernel for parallel electric and magnetic fields
can now be written down (using ${\cal E}=iE, c_0=i\mu$ to continue
to Minkowski space-time):
\be\label{4.7}
h^{(\beta)}(t|x,x)=\frac{B}{4\pi\sinh Bt}\frac{E}{4\pi\sin Et}\sum
^\infty_{n=-\infty}
e^{-n^2\beta^2 E/4\tan Et}\times e^{n\beta(Ex_1+\mu)}
\ee
where the integration over $k$ has eliminated
 all dependence on the
spatial coordinate $x_2$ and on the constant $c_3$ from the diagonal
local heat kernel.

%%%%%%%%%%%%%%%%%%%%%%%%%%%%%%%%%
\subsection{Arbitrary $B_1(x_2)$}
%%%%%%%%%%%%%%%%%%%%%%%%%%%%%%%%%

The factorization of the electric and magnetic sectors for parallel
$\vec {\cal E}$ and $\vec B$ fields can be further exploited. Let us replace
the potential $A_3=Bx_2+c_3$ above by an arbitrary function
$A_3(x_2)$ of $x_2$. Then the background magnetic field $B_1=\partial_2
A_3$  has an arbitrary dependence on $x_2$.
The modes (\ref{4.1}) still factorize, and eq. (\ref{4.3}) is
replaced by
\[[-\partial^2_2+(A_3(x_2)-k)^2]\Psi_{kn'}(x_2)=w^2_{kn'}\Psi_{kn'}
(x_2).\]
Now $\lambda^2_{mp}=2{\cal E}(n+1/2)+w^2_{kn'}$, with the (unknown)
eigenfunctions $\Psi_{kn'}(x_2)$ and spectrum $\{w_{kn'}^2\}$
determined by the mode equation just above. The heat kernel (\ref{4.7})
is replaced by
\[
h^{(\beta)}(t|x,x)=h^{(\beta)}_{2}(t|x_2,x_2)[E\;{\rm dependent}\;\;{\rm factor\ in\ eq. (\ref{4.7})}]
\]
with 
\[h_{2}(t|x_2,x_2)=\int dk\sum_{n'}e^{-tw^2_{kn'}}|\Psi_{kn'}
(x_2)|^2\]
in place of $h_{2}=B/4\pi\sinh Bt$. Obviously the Fourier series
structure of the heat kernel is preserved, even for arbitrary $B_1(x_2)$.

%%%%%%%%%%%%%%%%%%%%%%%%%%%%%%
\subsection{Cylindrical space}
%%%%%%%%%%%%%%%%%%%%%%%%%%%%%%

If spatial direction $x_3$ were compact -- say $0\leq x_3\leq L$ --
the conjugate momentum $k$ would be discrete: $k=r(2\pi/L)$ with $r=0,
\pm1,\pm2,...$. Then the integral (\ref{4.6}) would become a sum 
over $r$, exactly the same mode sum which leads to the Euclidean
version of the electric field factor in eq. (\ref{4.7}), with $L$ and $iB$
in place of $\beta$ and $E$. Thus the above compactification of
$x_3$ leads to the thermal heat kernel
\bear\label{4.9}
h^{(\beta)}(t|x,x)&=&\frac{B}{4\pi\sinh Bt}\sum_r e^{-r^2L^2B/\tanh Bt}
\times e^{irL(Bx_2+c_3)}\nonumber\\
&&\times\frac{E}{4\pi\sin Et}\sum_n e^{-n^2\beta^2E/\tan Et}\times
e^{n\beta(Ex_1+\mu)},
\ear
eq. (\ref{4.7}) being the $L=\infty$ limit of this. By compactifying the
spatial axis $x_3$, the gauge potential $A_3=x_2B+c_3$ has turned
into a compact local variable
$O\leq LA_3\leq 2\pi$,
very much as the compactification of Euclidean time leads to the
compactification of $\beta A^E_0$. This has nothing to do with the electric field 
and remains true at zero temperature and $E=0$.

Interesting mathematical physics is associated with the compactification
of $LA_3$; however, this lies beyond the scope of the present paper.
We mention some early literature (see e.g. refs. \cite{23, 24}) which
investigates the effect of $x_3$ compactification on a $T=0$ QFT.

%%%%%%%%%%%%%%%%%%%%%%%%%%%%%%%%%%%%%%%%%%%%%%%%%%%%
\section{Perpendicular Electric and Magnetic Fields}
\setcounter{equation}{0}
%%%%%%%%%%%%%%%%%%%%%%%%%%%%%%%%%%%%%%%%%%%%%%%%%%%%

Finally we work through the quite different problem of 
perpendicular background
$\vec E$ and $\vec B$-fields. For such a background the magnetic force on
moving charges has a component in the direction of the electrostatic
force on the same charge. This couples the electric and magnetic
sectors, eliminating the factorization observed for $\vec E
\parallel\vec B$ in the preceeding section. See refs. \cite{20}
-\cite{22} for 
the spinor version of this system (treated globally) and Schwinger
\cite{12} for the $T=0$ problem.

Choosing the Euclidean gauge potential $A^E_\mu(x)=
({\cal E}x_1+C_0,0,Bx_1+C_2,0)$ corresponding to background (Euclidean)
electric and magnetic fields $\vec{\cal E}=({\cal E},0,0)$ and
$\vec B=(0,0,B)$, respectively, the mode operator in eq. (\ref{2.5}) is
\bear\label{5.1}
-D^2&=&\left({\cal E} x_1+C_0-\frac{2\pi m}{\beta}\right)^2-
\partial^2_1+(-k_2+Bx_1+C_2)^2+k^2_3\nonumber\\
&=&-\partial^2_1+w^2(x_1-u)^2+v^2+k^2_3,\ear
where
\be\label{5.2}
w^2u={\cal E}\left(\frac{2\pi m}{\beta}-C_0
\right)+B(k_2-C_2),
\ee
\be\label{5.3}
w^2v^2=\left[B\left(\frac{2\pi m}{\beta}-C_0\right)
-{\cal E}(k_2-C_2)\right]^2,\ee
and $w^2={\cal E}^2+B^2$. We have included the constant term
in $A_2=Bx_1+C_2$  even though we know that $C_2$ cannot appear
in physical quantities, for it is of some interest to see how
the mathematics eliminates $C_2$. In eq. (\ref{5.1}) we have
assumed the modes (\ref{2.5}) (with $p=(n,k_2,k_3)$)
to be of the factorized form
\be\label{5.4}
\varphi_{mp}(\vec x)=\frac{1}{2\pi}e^{i(k_2x_2+k_3x_3)}
\varphi_{mn k_2}(x_1).
\ee
The eigenvalues $\lambda^2_{mp}$ of the operator (\ref{5.1}) are
then given by
\be\label{5.5}
\lambda_{mp}^2=2w(n+\frac{1}{2})+k^2_3+v^2
\ee
with $HO$ eigenfunctions $\varphi_{mnk_2}(x_1)=\varphi_n(x_1-u)$, where
$\varphi_n$ is given by
\be\label{HO}
\varphi_n(x)=2^{-n/2}\frac{1}{\sqrt{n!}}\left(
\frac{w}{\pi}\right)^{\frac{1}{4}}e^{-\frac{1}{2}{w}x^2}
H_n(\sqrt {w} x)\;,
\ee
The result of the calculation of the Euclidean thermal space-time heat kernel
(\ref{2.6}) is expedited by eq. (\ref{4.6}) with the substitution
$B \to w$. For the diagonal
heat kernel one finds
\bear\label{5.6}
h^{(\beta)}(t|x,x)&=&\frac{1}{\beta}\sum_m\frac{1}{4\pi^2}\int dk_2 dk_3 e^{-tk_3^2}e^{-tv^2}\nonumber\\
&&\times \left[\frac{w}{2\pi \sinh 2wt}\right]^{1/2}e^{-w\tanh wt(x_1-u)^2}\ear
with $u$ and $v^2$ given by eqs (\ref{5.2}), (\ref{5.3}). The change of
variable (which eliminates $C_2$)
\[wv={\cal E}(k_2-C_2)-B(\frac{2\pi m}{\beta}-C_0)\]
leads to
\bear\label{5.7}
h^{(\beta)}(t|x,x)&=&(4\pi t)^{-1/2}\left[\frac{w}{2\pi\sinh 2wt}\right]^{1/2}
\frac{w}{{\cal E}}\int^\infty_{-\infty}dv\ e^{-tv^2}\nonumber\\
&&\times \frac{1}{\beta}\sum^{\infty}_{m=-\infty} e^{-w\tanh wt(x_1-u)^2}\ear
with now
\[u=\frac{1}{{\cal E}}\left(\frac{2\pi m}{\beta}-C_0\right)
+\frac{B}{w{\cal E}}v.\]
The Matsubara sum is done with the help of a well-known theta function
identity 
\bear\label{5.8}
&&\frac{1}{\beta}\sum^\infty_{m=-\infty}e^{-w\tanh wt(x_1-u)^2}
={\cal E}\left[\frac{1}{4\pi w \tanh wt}\right]^{1/2}\nonumber\\
&&\times\sum^\infty_{n=-\infty}e^{-n^2{\cal E}\beta^2/4w\tanh wt}
e^{-in\beta({\cal E} x_1+C_0)}e^{in\beta Bv/w}\ear
Finally we employ
\[\frac{1}{2\pi}\int^\infty_{-\infty}dve^{-tv^2}e^{in\beta Bv/w}
=(4\pi t)^{-1/2}e^{-n^2\beta^2B^2/4w^2t}\]
to write the heat kernel (\ref{5.7}) in the form
\bear\label{5.9}
h^{(\beta)}(t|x,x)&=&(4\pi t)^{-1}\frac{w}{4\pi\sinh wt}\\
&\times& \sum^\infty_{m=-\infty}e^{-m^2\beta^2{\cal E}/4w\tanh wt}
e^{-m^2\beta^2B^2/4w^2t} e^{-im\beta({\cal E} x_1+C_0)}\;,
\nonumber
\ear
which has the expected  form (\ref{2.6}) with nonalternating sign.
Moreover, for $B\to 0$ or ${\cal E}\to 0$ this heat kernel has the
correct limits.

%%%%%%%%%%%%%%%%%%%%
\section{Conclusion}
%%%%%%%%%%%%%%%%%%%%

Our main result is that for a thermal charged matter field coupled to
a static electromagnetic background gauge potential $A_\mu(\vec x)$
the thermal plasma -- but not the virtual sea -- feels locally
the potential $A_0(\vec x)$ in addition to the gauge-invariant electric 
and magnetic fields $\vec E=-\vec\nabla A_0$ and
$\vec B=\vec\nabla\times \vec A$. This was discovered in the context of specific
calculations \cite{16,18} involving a constant background $\vec E$, with 
$\vec B=0$. Here we have explained the underlying general principles
and generalized the discussion to an arbitrary static potential $A_0(\vec x)$ 
(and hence also an arbitrary static electric field) and an arbitrary
static magnetic field $\vec B(\vec x)$. For reasons of gauge invariance the Euclidean gauge potential
$A^E_0(\vec x)$ is a local compact variable in any local function
describing the many-body quantum system. This function therefore
has a Fourier cosine series expansion in $\beta A^E_0(\vec x)$,
in which the term independent of $A^E_0$ represents the virtual sea
contribution. Continued to Minkowski spacetime, this series becomes
a hyperbolic cosine expansion in the Minkowski potential $\beta A_0(\vec x)$,
displaying the chemical-potential-like role of a constant background voltage
for the charged thermal field.

In sections 4 and 5 we then extended our previous explicit scalar field
calculation with $\vec B=0$ to the two most important backgrounds
with uniform $\vec E$ and $\vec B$: namely $\vec E\parallel \vec B$
and $\vec
E\perp\vec B$. Completely explicit Fourier series were obtained
for the thermal heat kernels of these systems, thereby providing
additional, more complex examples of the general theory.
For brevity we have not included (although one easily could)
effective Lagrangians, energy momentum tensors and the like in these
examples. Our goal has been to provide new insight into the local
aspects of thermal matter fields coupled to static electromagnetic
backgrounds. We hope to present more complete results
for interesting systems at a later time.

\end{document}